\documentclass{statsoc}

\usepackage[english,activeacute]{babel}
\usepackage[a4paper]{geometry}
\usepackage{graphicx}
\usepackage[textwidth=8em,textsize=small]{todonotes}
\usepackage{amsmath}
\usepackage{amssymb}
\usepackage{natbib}
\usepackage{url}
\usepackage{enumerate}
\usepackage{multicol}
\usepackage{multirow}
\usepackage[T1]{fontenc}
\usepackage{float}
\usepackage{subfigure}
\usepackage{rotating}
\usepackage{ulem}

\restylefloat{table}

\def\exp#1{{\rm exp}{#1}}
\def\frac#1#2{{{#1}\over{#2}}}
\def\binom#1#2{{{#1}\choose{#2}}}

\def\le{\left}
\def\ri{\right}

\DeclareMathOperator*{\expit}{expit}

\newcommand\simiid{\mathrel{\overset{\makebox[0pt]{\mbox{\normalfont\tiny\sffamily iid}}}{\sim}}}
\newcommand\simind{\mathrel{\overset{\makebox[0pt]{\mbox{\normalfont\tiny\sffamily ind}}}{\sim}}}

\newcommand{\ind}[1]{\mathbb{I}\left\{ #1 \right\}}
\newcommand{\pr}[1]{\mathbb{P}\text{r}\left[#1\right]}
\newcommand{\expec}[1]{\mathbb{E}\left[#1\right]}

\newcommand{\var}[1]{\mathbb{V}\text{ar}\left[#1\right]}
\newcommand{\sd}[1]{\text{SD}\left[#1\right]}

\newcommand{\ex}[1]{\exp{ \left\{ #1 \right\}}}

\def\M{\mathbf{M}}

\def\P{\mathbf{P}}

\def\rv{\boldsymbol{r}}

\def\uv{\boldsymbol{u}}

\def\wv{\boldsymbol{w}}
\def\xv{\boldsymbol{x}}
\def\Y{\mathbf{Y}}

\def\be{\beta}

\def\del{\delta}
\def\eps{\epsilon}

\def\te{\theta}\def\tev{\boldsymbol{\theta}}
\def\vtev{\boldsymbol{\vartheta}}

\def\ka{\kappa}

\def\sig{\sigma}

\def\ome{\omega}

\def\xiv{\boldsymbol{\xi}}
\def\psiv{\boldsymbol{\psi}}

\def\piv{\boldsymbol{\pi}}
\def\phiv{\boldsymbol{\phi}}
\def\Ga{\Gamma}

\def\UPS{\mathbf{\Upsilon}}

\def\Cat{\small{\mathsf{Cat}}}

\def\DP{\small{\mathsf{DP}}}
\def\Ber{\small{\mathsf{Ber}}}

\def\Nor{\small{\mathsf{N}}}

\def\Bet{\small{\mathsf{Beta}}}

\def\Gamd{\small{\mathsf{Gam}}}

\def\data{\text{data}}

\def\rest{\text{rest}}

\def\rest{\text{rest}}

\def\zerov{\boldsymbol{0}}

\title[De-duplication in the Presence of Relational Data]{A Bayesian Approach for De-duplication in the Presence of Relational Data}
\author[Juan Sosa]{Juan Sosa}
\address{Universidad Nacional de Colombia}
\email{jcsosam@unal.edu.co}
\author{Abel Rodr\'iguez}
\address{University of Washington}
\email{abelrod@uw.edu}

\begin{document}

\begin{abstract}
In this paper, we study the impact of combining profile and network data in a de-duplication setting. We also assess the influence of a range of prior distributions on the linkage structure. Furthermore, we explore stochastic gradient Hamiltonian Monte Carlo methods as a faster alternative to obtain samples from the posterior distribution for network parameters. Our methodology is evaluated using the RLdata500 data, which is a popular dataset in the record linkage literature.

\vspace{6pt}

\hspace{6pt} \textit{Keywords.}  Allelic Partitions; Microclustering; Network data; Latent space models; Record Linkage.
\end{abstract}

\section{Introduction}

In database management, record de-duplication aims to identify multiple records that correspond to the same individual. This process can be treated as a clustering problem, in which each latent entity is associated with one or more noisy database records.  From a model-based perspective, popular choices for clustering include finite mixture models and Dirichlet/Pitman-Yor process mixture models (\citealp{muller-2013}, \citealp{casella-2014}, \citealp{miller-2016-mixture}). Although these alternatives have proven to be successful in all sorts of applications, they are not realistic for de-duplication problems.

Some approaches for de-duplication have been considered during the past few years. \cite{domingos-2004} treat the problem of de-duplication within one file through an uni-partite graph, allowing information to propagate from one candidate match to another via the attributes they have in common. \citet{sadinle-2013} and \citet{sadinle-2014} look for duplicate records  partitioning the data file into groups of coreferent records. They present an approach that targets this partition of the file as the parameter of interest, thereby ensuring transitive decisions. The work of \citet{steorts2015entity} and \cite{steorts2016bayesian} also permit de-duplication while handling multiple files simultaneously.
Other recent contributions in this area that have proven to be very useful are \cite{enamorado2020probabilistic}, \cite{tancredi2020unified}, \cite{marchant2021d}, and \cite{aleshinguendel2021multifile}.

Unlike models exhibiting infinitely exchangeable clustering features, models specifically conceived for entity resolution (ER) need to generate small clusters with a minor number of records, no matter how large the database is. Specifically, we require clusters whose sizes grow sublinearly with the total number of records in order to accurately identify the latent entity underlying each observed record (\citealp{miller-2015}, \citealp{betancourt-2016}, \citealp{betancourt2020random}, \citealp{betancourt2020prior}).

On the other hand, findings in \cite{sosa2018record} show that network data can substantially improve merging online social networks (OSNs). Hence, it make sense that network data can be also useful in other ER tasks such as de-duplication. This might be useful, for example, in identifying covert users in a social network, which might have multiple profiles linking to the same groups of individuals.
Thus, our goal in this manuscript is three-fold.  First, we extend the model in \cite{sosa2018record} from OSNs matching to handle de-duplication tasks.  Second, we examine a range of priors on the linkage structure (cluster assignments), and then assess their influence on the posterior linkage. And finally, we also explore stochastic gradient Hamiltonian Monte Carlo methods \citep{chen-2014} as a faster way to obtain samples from the posterior distribution for network parameters.

The remainder of this article is organized as follows: Section \ref{sec_an_entity_resolution_model} introduces a model for de-duplication handling both attribute and relational data; there, we discuss in detail every aspect of the model including prior specification and computation. Section \ref{sec_microclustering} examines in detail the concept of microclustering. Section \ref{sec_prior_specification_on_the_linkage_structure} presents a number of prior distributions on the linkage structure. Section \ref{sec_RL500_dataset} compares the performance of the resulting procedures using the RLdata500 data, a popular dataset in the record linkage literature. Section \ref{sec_sensitivity_analysis} explores the robustness of the results to the prior specification and the structural features of the network information.
Section \ref{sec_fast_computation} presents a faster way to draw samples for the network parameters based on stochastic gradient methods. Lastly, we discuss our findings and directions for future work in Section \ref{sec_discussion}.

\section{A de-duplication model incorporating relational data}\label{sec_an_entity_resolution_model}

We rely on the formulation provided in \cite{steorts2016bayesian} and \cite{sosa2018record} for $J=1$ file. Specifically, we have a single file with $I$ records, each containing $L$ fields in common, for which both profile data $\P=[p_{i,\ell}]$ and network data $\Y=[y_{i,i'}]$ are available in order to uncover multiple records corresponding to the same latent identity.

\subsection{Model formulation}

We model both sources of information independently given the linkage structure $\xiv = (\xi_1,\ldots,\xi_I)$, which defines a partition $\mathcal{C}_{\xiv}$ on $\{1,\ldots,I\}$. Entries in $\xiv$ are labeled consecutively from 1 to $N$. See Section 3 in \cite{sosa2018record} for more details about notation and the nature of the problem.

Accordingly, we model the relational data through a latent distance model \citep{hoff-2002} of the form \begin{equation}\label{eq_ER_model_link_data}
	y_{i,i'}\mid \beta, \uv_{\xi_{i}},\uv_{\xi_{i'}} \simind \Ber\le(\expit\le(\be - \|\uv_{\xi_{i}} - \uv_{\xi_{i'}}\|\ri)\ri),
\end{equation}
where each $\uv_n$ is embedded in a $K$-dimensional social space. Then, the attribute data are modeled according to the status of field-specific distortion indicators $w_{i,\ell}$ through
\begin{equation}\label{eq_ER_model_profile_data}
	p_{i,\ell}\mid\pi_{\xi_{i},\ell},w_{i,\ell},\vtev_\ell\simind
	\left\{
	\begin{array}{ll}
		\del_{\pi_{\xi_{i},\ell}}, & \hbox{$w_{i, \ell} = 0$;} \\
		\Cat(\vtev_\ell), & \hbox{$w_{i, \ell}=1$,}
	\end{array}
	\right.\\
\end{equation}
where $\vtev_\ell$ is an $M_\ell$-dimensional vector of multinomial probabilities.
The rest of the model, but the prior specification on $\xiv$, is given exactly as in \citet[Sec. 3 and 4]{sosa2018record}. Lastly, we devote Section \ref{sec_prior_specification_on_the_linkage_structure} in this document to discuss several prior formulations for $\xiv$.

\subsection{Hyperparameter elicitation}\label{sec_elicitation}

Following the same line of thought given in \cite{krivitsky2008fitting}, we let the network hyperparameters take the values $\ome=100$, $a_\sig = 2 + 0.5^{-2}$, and $b_\sig=(a_\sig-1)\,\tfrac{\sqrt{I}}{\sqrt{I} - 2}\, \tfrac{\pi^{K/2}}{ \Gamma(K/2 + 1)}\, I^{2/K}$. On the other hand, for the profile hyperparameters we follow very closely \cite{steorts2015entity} and set $\alpha_{\ell,m}=1$, $a_{\ell} = a = 1$, and $b_\ell = b = 99$.

\subsection{Computation}\label{sec_computation}

Computation for this model can still be achieved via Markov chain Monte Carlo (MCMC) algorithms \citep{gamerman2006markov}. Hence, the full set of parameters in this case is
$$
\UPS = \left( \xiv, \phiv, \uv_{1},\ldots,\uv_{N}, \beta, \sig, \piv_1, \ldots,\piv_N, \vtev_1,\ldots,\vtev_L, \wv, \psiv \right)\,
$$
where $\phiv$ includes those parameters in the prior distribution of $\xiv$. Full conditional distributions are available in closed form for all the profile parameters.  As far as the network parameters is concerned, random walk Metropolis-Hastings steps can be used. The Appendix provides details to sample all the model parameters.  Recall that the main inference goal is to make inferences about $\xiv$ by drawing samples $\xiv^{(1)},\ldots,\xiv^{(S)}$ from the posterior distribution $p(\UPS\mid\data)$ and then getting a point estimate of the overall linkage structure.

\section{Microclustering}\label{sec_microclustering}

Finite mixture models and Dirichlet/Pitman-Yor process mixture models
are widely used in many clustering applications \citep{miller-2016-mixture}. These models generate cluster sizes that grow linearly with the number of records $I$, i.e., for all $n$, $\tfrac{1}{I}\sum_{i}\ind{\xi_i=n}\xrightarrow{\text{a.s.}} \pr{\xi_i=n}$ when $I\rightarrow\infty$. Such a property is unappealing to address de-duplication problems because we need to generate a large number clusters with a negligible number of records (mostly singletons and pairs).

In order to formulate more realistic models for de-duplication, \cite{miller-2015} introduce the concept of microclustering, in which the model is required to produce clusters whose sizes grow sublinearly with $I$. Formally, a model exhibits the microclustering property if $\tfrac{M}{I}\xrightarrow{\text{p}}0$ as $I\rightarrow\infty$, where $M=\max\le\{|C_n|:C_n\in \mathcal{C}_{\xiv}\ri\}$ is the size of the largest cluster in $\mathcal{C}_{\xiv}$. No mixture model can exhibit the microclustering property, unless its parameters are allowed to vary with $I$ \citep{betancourt-2016}.

\cite{miller-2015} show that in order to obtain nontrivial models exhibiting the microclustering property, we must sacrifice either finite exchangeability or projectivity. We follow \cite{betancourt-2016} in that regard and enforce the former since sacrificing projectivity is less restrictive in the context of ER. As a consequence, inference on $\xiv$ will not depend on the order of the data, but the implied joint distribution over a subset of records will not be the same as the joint distribution obtained by modeling the subset directly. Previous work of \cite{wallach-2010} sacrifices exchangeability instead.

\section{Prior specification on the linkage structure}\label{sec_prior_specification_on_the_linkage_structure}

\subsection{Kolchin partition priors}

The Kolchin partition priors (KPPs) are originally introduced in \cite{betancourt-2016} as a way to enforce the microclustering property. This approach consists in placing a prior on the number of clusters, $N \sim \ka$, and then, given $N$, the cluster sizes $S_1,\ldots,S_N$ are modeled directly $S_1,\ldots,S_N\mid N \simiid \mu$. Here $\ka$ and $\mu$ are probability distributions over  $\mathbb{N}=\{1,2,\ldots\}$. In this way, given $I = \sum_{n=1}^N S_n$, it is straightforward to generate a set of cluster assignments $\xiv=(\xi_1,\ldots,\xi_I)$, which in turn induces a random partition $\mathcal{C}_{\xiv} = \{C_1,\ldots,C_N\}$, by drawing a vector uniformly at random from the set of permutations of $1,\ldots,1$ ($S_1$ times), $\ldots$, $N,\ldots,N$ ($S_N$ times). Hence, conditioning on $I$ (the total number of records is usually observed), it can be shown that the probability of any given partition is
$$
\pr{\mathcal{C}_{\xiv}\mid I} \propto |\mathcal{C}_{\xiv}|\,\kappa\le(|\mathcal{C}_{\xiv}|\ri)\,\prod_{n=1}^N |C_n|!\,\mu\le(|C_n|\ri),
$$
where $|\cdot|$ denotes the cardinality of a set. We discuss below two particular choices of $\ka$ and $\mu$ that have proven to exhibit the microclustering property, and adopt them as a baseline in Section \ref{sec_RL500_dataset}. We remit the reader to \cite{miller-2015}, \cite{betancourt-2016}, and \cite{betancourt2020random} for details about computation and prior elicitation.

The Negative Binomial--Negative Binomial Prior (NBNBP)
assumes that both $\ka$ and $\mu$ are negative binomial distributions (truncated to $\mathbb{N}$) with parameters $a$ and $q$ and $\eta$ and $\theta$, respectively. Here, $a > 0$ and $q \in (0, 1)$ are fixed hyperparameters, while $\eta>0$ and $\theta\in (0,1)$ are distributed as $\eta \sim \Gamd(a_\eta, b_\eta)$ and $\theta \sim \Bet(a_\theta, b_\theta)$ for fixed hyper-parameters $a_\eta, b_\eta, a_\theta,b_\theta$. When evaluating the performance of this prior, we follow the authors and set $a$ and $q$ in a way that $\expec{N} = \sqrt{\var{N}} = \tfrac{I}{2}$, $a_\eta = b_\eta = 1$, and $a_\theta = b_\theta = 2$.

The Negative Binomial--Dirichlet Prior (NBDP)
still assumes that $\ka$ is a negative binomial distribution (truncated to $\mathbb{N}$) with parameters $a$ and $q$, but this time $\mu\sim\DP(\alpha,\mu^0)$. Here, $a$ and $q$ are once again fixed hyperparameters, $\alpha$ is a fixed concentration parameter and $\mu^0$ is a fixed base measure with $\sum_{m=1}^\infty\mu^0(m)=1$ and $\mu^0(m)\geq 0$, for all $m$. The parameters $a$ and $q$ are set as before, while $\alpha = 1$ and $\mu^0$ is set to be a geometric distribution over $\mathbb{N}$ with parameter 0.5.

\subsection{Allelic partition priors}

Here we consider a class of prior distributions on the cluster assignments $\xiv$ based on allelic partitions \citep{crane-2016}. Let $\mathcal{C}_{\xiv} = \{C_1,\ldots,C_N\}$ be the partition implicitly represented by $\xiv$, and let $\rv=(r_1,\ldots,r_I)$ be the allelic partition induced by $\mathcal{C}_{\xiv}$, where $r_i$ denotes the number of clusters of size $i$ in $\mathcal{C}_{\xiv}$. 
Assuming that partitions corresponding to the same allelic partition occur with the same probability, we can generate a random partition by first drawing an allelic partition and then selecting uniformly among partitions for which that specific allelic partition holds. This simple reasoning allow us to write 
$$
p(\xiv) = \frac{1}{I!} \prod_{i=1}^I i!^{r_i}\,r_i! \times p(\rv)\,,
$$
which fully determines an exchangeable partition probability function. Thus, we just need to place a distribution on $\rv$ in order to complete the prior specification. We discuss below two instances that can be framed in the context of allelic partitions. We remit the reader to \citep{betancourt2020prior} for details about computation and prior elicitation.

We consider the allelic binomial prior (ABP, \citealp{betancourt2020prior}), setting the maximum cluster size to $M=2$, which leads to the allelic partition $\rv=(I - 2r_2,r_2,0, 0, \ldots,0)$, and then letting $r_2\sim\textsf{Beta-Bin}(a_2,b_2)$. This approach guarantees that the microclustering property holds, because the value of $M$ is being handled directly. We let $a_{2}=\tfrac{\rho-\gamma^2}{(1+\rho)\gamma^2}$ and $b_{2}=a_{2}\,\rho$, with $\rho=(1-\pi)/\pi$, where $\pi=0.8$ is the prior probability of expecting a singleton, and $\gamma=0.5$ is the corresponding coefficient of variation.

Finally, another popular alternative that does not satisfy the microclustering property but is convenient for practical reasons, is the Ewens-Pitman Prior (EPP, \citealp{mccullagh-2006}). The probability mass function for the EPP is given by $p(\xiv\mid\theta) = \tfrac{\Ga(\te)}{\Ga(I+\te)}\,\theta^N \prod_{n=1}^N\Ga(S_n)$, with $\theta \sim\Gamd(a_\theta,b_\theta)$. The parameters $a_\theta$ and $b_\theta$ are carefully chosen in order to match the prior beliefs given in the ABP.

\section{Evaluation}\label{sec_RL500_dataset}

We investigate the impact of including relational data in the de-duplication process as well as the performance of the ABP compared to other existing priors. To this end, we consider the RLdata500 dataset from the RecordLinkage package \citep{RL-package} in R, which has been considered by many authors to test their methodologies, including  \cite{christen-2009-accurate}, \cite{christen-2013-flexible}, \cite{steorts-2014-comparison}, \cite{steorts2015entity}, and \cite{tancredi2020unified}. This is a syntectic dataset with $I=500$ records, 50 of which are duplicates. Each record has associated with it seven fields, namely, name's first component, name's second component, last name's first component, last name's second component, year of birth, month of birth, and day of birth. We only consider the last three fields (categorical fields) for illustrative purposes. The ground truth (true cluster assignments) is also available.

We augment this dataset by generating social ties between records following the latent distance model \eqref{eq_ER_model_link_data}, where $u_{n,k}\mid\sig^2\simiid \Nor(0,\sig^2)$ and $\xiv$ corresponds to the true linkage structure in the dataset. We consider two scenarios (see Table \ref{tab_nets_characteristics}), which allows us to study how structural features influence the de-duplication process.

\begin{table}[!ht]
	\centering
	\begin{tabular}{c|cccccc} \hline
		Scenario & $\beta$ & $\sig^2$ &$K$ & Transitivity & Assortativity  & Density  \\ \hline
		Scenario 1  &  10 & 178 & 2     & 0.576        & 0.680          & 0.126 \\
		Scenario 2  &  10 & 278 & 2    & 0.562        & 0.754          & 0.082 \\
		\hline
	\end{tabular}
	\caption{\label{tab_nets_characteristics}\footnotesize{Features of the network data.}}
\end{table}

\begin{figure}[!t]
	\centering
	\subfigure[UP]    {\includegraphics[scale=.36,]{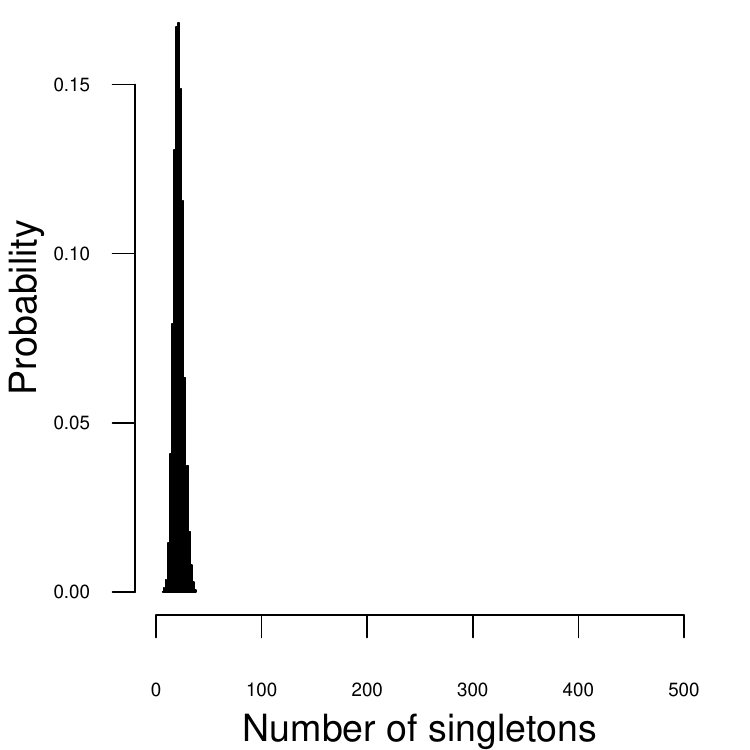}}
	\subfigure[ABP1]  {\includegraphics[scale=.36,]{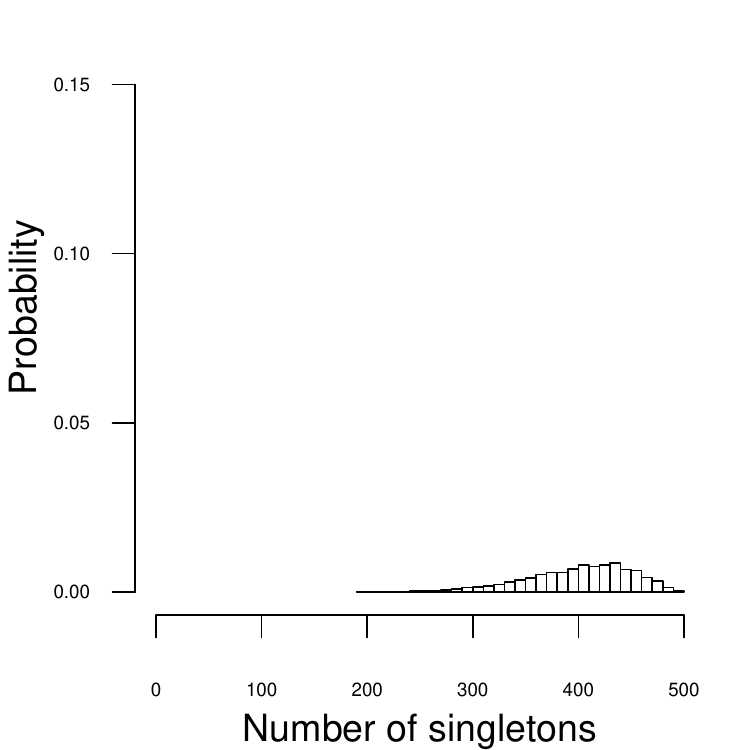}}
	\subfigure[ABP2]  {\includegraphics[scale=.36,]{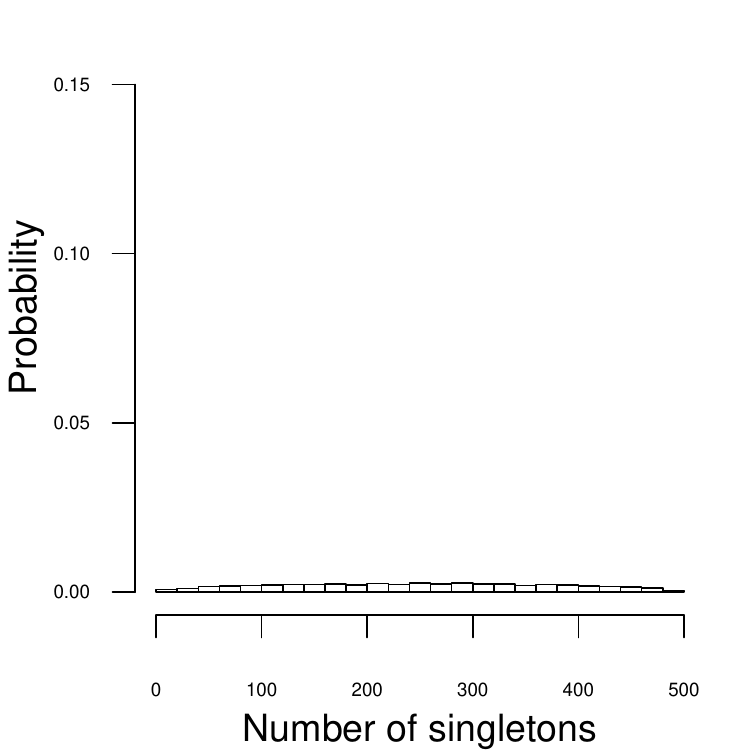}} \\
	\subfigure[EPP2] {\includegraphics[scale=.36,]{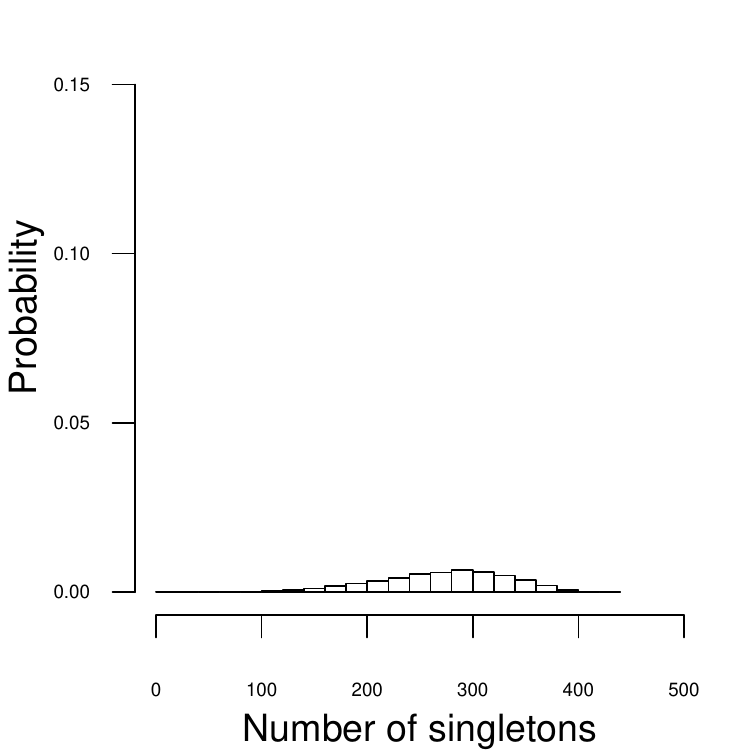}}
	\subfigure[NBNBP]{\includegraphics[scale=.36,]{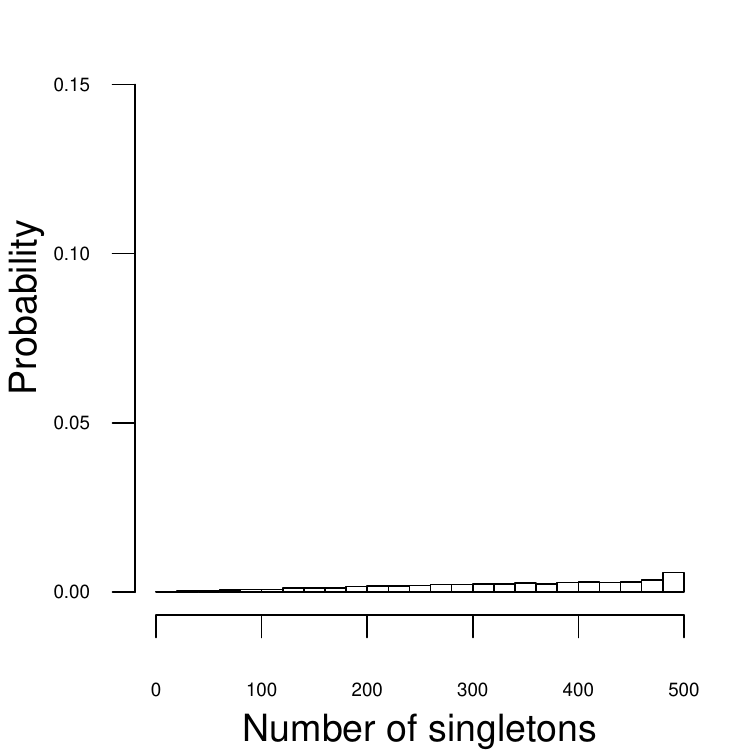}}
	\subfigure[NBDP] {\includegraphics[scale=.36,]{prior_sim_NBNBP.pdf}}
	\caption{\footnotesize{Prior distribution of the number of singleton clusters.}}
	\label{fig_prior_sim}
\end{figure}

We fit our de-duplication model using just profile data as well as using both profile and network data with $K=2$. We also implement each prior specification given in Section \ref{sec_prior_specification_on_the_linkage_structure}, along with an uniform prior (UP) as in \cite{steorts2016bayesian}. In particular, we calibrate the ABP, in such a way that 80\% and 50\% of clusters are a priori singleton clusters with a 0.5 coefficient of variation for $M=2$ (ABP1 and ABP2, respectively). The EPP is aslo calibrated in a similar fashion. We also calibrate the ABP around 80\% of singleton clusters with a 0.5 coefficient of variation for $M=3$ (ABP3). Histograms of the number of singleton clusters for some of these prior distributions are shown in Figure \ref{fig_prior_sim}. Lastly, we run the Gibbs sampler described in Appendix \ref{ap_RL1} based on 100,000 samples obtained after a burn-in period of 500,000 iterations. In addition, the clustering methodology proposed by \citet{lau-2007} was used to obtain a point estimate of the posterior linkage structure.

\begin{table}[!t]
	\centering
	\begin{tabular}{lccccc}
		\hline
		Prior & Recall & Precision & $\text{F}_1$ & $\expec{N\mid\text{data}}$ & $\sd{N\mid\text{data}}$ \\
		\hline
		\multicolumn{6}{c}{Profile data} \\
		\hline
		UP    & 0.62 & 0.45 & 0.52 & 264.78 & 2.63 \\
		\textbf{ABP1}  & \textbf{0.58} & \textbf{0.88} & \textbf{0.70} & \textbf{467.86} & \textbf{2.83} \\
		\textbf{ABP2}  & \textbf{0.58} & \textbf{0.88} & \textbf{0.70} & \textbf{467.97} & \textbf{2.85} \\
		\textbf{ABP3}  & \textbf{0.58} & \textbf{0.88} & \textbf{0.70} & \textbf{468.43} & \textbf{3.85} \\
		EPP1  & 0.02 & 0.50 & 0.04 & 497.97 & 0.74 \\
		EPP2  & 0.06 & 1.00 & 0.11 & 492.06 & 1.87 \\
		NBDP  & 0.58 & 0.88 & 0.70 & 469.19 & 2.66 \\
		NBNBP & 0.58 & 0.88 & 0.70 & 467.89 & 2.62 \\
		\hline
		\multicolumn{6}{c}{Profile and network data (Scenario 1)} \\
		\hline
		UP    & 1.00 & 0.32 & 0.49 & 344.12 & 1.56 \\
		\textbf{ABP1}  & \textbf{0.94} & \textbf{0.94} & \textbf{0.94} & \textbf{450.20} & \textbf{2.28} \\
		\textbf{ABP2}  & \textbf{0.94} & \textbf{0.92} & \textbf{0.93} & \textbf{450.35} & \textbf{1.12} \\
		\textbf{ABP3}  & \textbf{0.94} & \textbf{0.94} & \textbf{0.94} & \textbf{447.81} & \textbf{0.77} \\
		EPP1  & 0.84 & 0.81 & 0.82 & 449.38 & 2.19 \\
		EPP2  & 0.80 & 0.85 & 0.82 & 450.74 & 3.19 \\
		NBDP  & 0.94 & 0.71 & 0.81 & 445.66 & 2.46 \\
		NBNBP & 0.92 & 0.82 & 0.87 & 441.45 & 1.47 \\
		\hline
		\multicolumn{6}{c}{Profile and network data (Scenario 2)} \\
		\hline
		UP    & 0.94 & 0.31 & 0.47 & 346.34 & 1.91 \\
		\textbf{ABP1}  & \textbf{0.90} & \textbf{0.92} & \textbf{0.91} & \textbf{450.32} & \textbf{1.23} \\
		\textbf{ABP2}  & \textbf{0.90} & \textbf{0.94} & \textbf{0.92} & \textbf{451.85} & \textbf{1.44} \\
		\textbf{ABP3}  & \textbf{0.94} & \textbf{0.92} & \textbf{0.93} & \textbf{448.28} & \textbf{0.45} \\
		EPP1  & 0.76 & 0.70 & 0.73 & 447.20 & 4.53 \\
		EPP2  & 0.84 & 0.78 & 0.81 & 448.85 & 2.67 \\
		NBDP  & 0.92 & 0.82 & 0.87 & 441.34 & 2.29 \\
		NBNBP & 0.90 & 0.75 & 0.82 & 443.50 & 1.69 \\
		\hline
	\end{tabular}
	\caption{\label{tab_PM_PNM_results}\footnotesize{Performance assessment and summary statistics for each prior distribution using just profile data and also using both profile and network data. Only categorical fields are considered.}}
\end{table}

We report the results of our experiments in table \ref{tab_PM_PNM_results}.
When the model is fitted using only profile data, the recall of the procedure is relatively. There seems to be no difference between our prior and the KPPs in this setting. On the other hand, notice that the EPP's behavior is particularly poor; this fact suggests that satisfying the microclustering property is crucial, specially when only profile information is available and the number of fields is small. Even though the UP's recall seems higher, its precision is substantially low. In general, the population size is being overestimated; this is not the case for the UP because it has such a strong pull towards a small number of singletons as shown in Figure \ref{fig_prior_sim}.

As expected, including network data substantially improves the accuracy of the posterior linkage as well as the estimate of the population size; specially in cases like these, where profile data is not abundant. In general, every prior seems to favor a fair estimate of the population size, except the UP. On the other hand, looking at the $\text{F}_1$ score, the models based on our prior clearly outperform the rest. Interestingly, there is not much difference in performance between ABP1 and ABP2. Not surprisingly, those priors that do not satisfy the microclustering property perform worse than those that do. Notice also that the ABP produces similar results for both $M=2$ and $M=3$. Lastly, it seems to be the case that accuracy values tend to decrease a little when the network data is less dense. This feature is more evident for the EPP.

\section{Sensitivity analysis}\label{sec_sensitivity_analysis}

We fitted our de-duplication model making specific choices for several quantities. Specifically, we chose $\psi_\ell\simiid \Bet(a_\ell,b_\ell)$, with $a_\ell = a = 1$ and $b_\ell = b =99$. Here we consider the effect of varying the values of $a$ and $b$ on the posterior linkage and the estimate of the population size. To this end, we fit our model again using both profile and network data along with the ABP2 as a prior distribution for the linkage structure.

We explore several cases to assess the robustness of our model to the choice of $a$ and $b$. First, we fix the prior mean of each distortion probability at $a/(a+b) = 0.002$ (instead of 0.01) and vary $a$ and $b$ proportionally, which decreases the variance of the prior distribution. Then, we consider the effect of varying the prior mean $a/(a + b)$ while holding $a + b$ fixed at either $a + b = 100$ or $a + b = 10$. Results are shown in Table \ref{tab_sensitivity_results}.

We see that these results are fairly consistent to those presented in the second panel of Table \ref{tab_PM_PNM_results}, although there is a non-negligible improvement when $a=0.1$ and $b=49.9$; such a setting makes both recall and precision almost perfect as well as the estimate of the population size. On the other hand, precision tends to decrease when the prior variance of the distortion probabilities increases, e.g., $a=10$ and $b=90$, and also $a=1$ and $b=9$; prior specifications of this kind also lead to an underestimate of the population size. These findings suggest that our approach is quite robust to the prior specification of the distortion probabilities.

\begin{table}[ht]
	\centering
	\begin{tabular}{ccccccc}
		\hline
		$a$ & $b$ & Recall & Precision & $\text{F}_1$ & $\expec{N\mid\text{data}}$ & $\sd{N\mid\text{data}}$ \\
		\hline
		\multicolumn{7}{c}{$a/(a+b) = 0.002$} \\
		\hline
		0.004 & 1.996  & 0.94 & 0.90 & 0.92 & 447.48 & 0.57 \\
		0.010 & 4.990  & 0.94 & 0.87 & 0.90 & 445.79 & 0.61 \\
		0.020 & 9.980  & 0.94 & 0.84 & 0.89 & 442.42 & 1.02 \\
		0.040 & 19.960 & 0.96 & 0.89 & 0.92 & 445.39 & 1.62 \\
		\textbf{0.100} & \textbf{49.900} & \textbf{0.96} & \textbf{0.96} & \textbf{0.96} & \textbf{449.26} & \textbf{0.57} \\
		0.200 & 99.800 & 0.98 & 0.91 & 0.94 & 446.46 & 0.50 \\
		\hline
		\multicolumn{7}{c}{$a+b=100$} \\
		\hline
		0.030  & 99.970 & 0.96 & 0.92 & 0.94 & 448.02 & 1.07 \\
		0.100  & 99.900 & 0.92 & 0.84 & 0.88 & 444.57 & 1.75 \\
		0.300  & 99.700 & 0.96 & 0.91 & 0.93 & 446.56 & 1.83 \\
		1.000  & 99.000 & 0.94 & 0.92 & 0.93 & 450.35 & 1.12 \\
		3.000  & 97.000 & 0.96 & 0.89 & 0.92 & 445.51 & 0.86 \\
		10.000 & 90.000 & 0.94 & 0.82 & 0.88 & 441.76 & 0.85 \\
		\hline
		\multicolumn{7}{c}{$a+b=10$} \\
		\hline
		0.003 & 9.997 & 0.96 & 0.91 & 0.93 & 446.56 & 0.69 \\
		0.010 & 9.990 & 0.96 & 0.89 & 0.92 & 445.20 & 0.62 \\
		0.030 & 9.970 & 0.92 & 0.92 & 0.92 & 449.39 & 1.08 \\
		0.100 & 9.900 & 0.94 & 0.87 & 0.90 & 445.37 & 0.73 \\
		0.300 & 9.700 & 0.92 & 0.92 & 0.92 & 449.17 & 0.37 \\
		1.000 & 9.000 & 0.96 & 0.81 & 0.88 & 440.54 & 2.55 \\
		\hline
	\end{tabular}
	\caption{\label{tab_sensitivity_results}\footnotesize{Performance assessment and summary statistics for the ABP2 using both profile and network data (Scenario 1). Several values of $a$ and $b$ have been considered.}}
\end{table}

\section{An alternative way to draw samples for the network parameters }\label{sec_fast_computation}

Suppose we want to generate samples from the posterior distribution of $\tev$ given a set of independent observations $\xv \in \mathrm{D}$, $p(\tev\mid\mathrm{D}) \propto \ex{-U(\tev)}$, where the potential energy function $U$ is given by $U(\tev) = -\sum_{\xv\in\mathrm{D}} \log p(\xv\mid\tev) - \log p(\tev)$. A Hamiltonian Monte Carlo (HMC) algorithm introduces a set of auxiliary variables $\rv$ and draws samples from the joint distribution $p(\tev,\rv)\propto\ex{-U(\tev)-\tfrac12 \rv^T \M \rv}$ by simulating from a Hamiltonian system, where $\M$ is a mass matrix usually defined as the identity matrix. If we simply discard the resulting $\rv$ samples, the $\tev$ samples have marginal distribution $p(\tev\mid \mathrm{D})$. See \cite{neal-2011} for details.

Now, along the lines of \cite{chen-2014}, instead of computing the gradient $\nabla U(\tev)$ using the entire dataset $\mathrm{D}$, the stochastic gradient HMC (SGHMC) considers a noisy estimate based on a minibatch $\tilde{\mathrm{D}}$ sampled uniformly at random from $\mathrm{D}$:
$$
\nabla\tilde{U}(\theta) = -\frac{|\mathrm{D}|}{|\tilde{\mathrm{D}}|} \sum_{\xv\in\tilde{\mathrm{D}}} \log p(\xv\mid\tev) - \log p(\tev),
$$
where $|\cdot|$ denotes the cardinality of a set. Clearly, we want minibatches to be small in order to obtain a significant reduction in the computational cost of $\nabla U(\tev)$. Details about the SGHMC are provided in Appendix \ref{ap_RL1}.

We want to compare a random-walk (RW) and a SGHMC in terms of accuracy and computational cost using the data provided in Section \ref{sec_RL500_dataset}. Once again we fit our de-duplication model using both profile and network data, and the ABP2 as a prior distribution for the linkage structure. To do so, we follow the algorithm outlined in the Appendix using both a RW and a SGHMC to sample from the conditional distribution of $\beta$ and each $\uv_n$. The RW adaptively finds the value of the tunning parameter in order to automatically find a good proposal distribution. Regarding the SGHMC, we set the mass matrix $\M$ to the identity matrix; after some experimentation, we decided to make the scaling factor $\eps=0.001$ and the number of leapfrogs steps $L=5$. Such values provide reasonable acceptance rates in this case. Lastly, minibatches are chosen by sampling uniformly at random 20\% of the corresponding data points. We run both algorithms based on 100,000 samples obtained after a burn-in period of 500,000 iterations.

Table \ref{tab_results_HMC} shows the corresponding results. We see that the SGHMC provides sensible levels of accuracy in comparison with the RW. In particular, both approaches yield to extremely good recall values. Even though we loose some precision with the SGHMC, we reduce the computation time around 43\%. These results are comparable with those in Table \ref{tab_PM_PNM_results}, where fitting the model using other prior distributions such as the EPP and the KPPs produces similar levels of accuracy.

\begin{table}[ht]
	\centering
	\begin{tabular}{c|cccccc}
		\hline
		Algorithm & Recall & Precision & $\text{F}_1$ & $\expec{N\mid\text{data}}$ & $\sd{N\mid\text{data}}$ & Time sec/100 \\
		\hline
		RW    & 0.94 & 0.92 & 0.93 & 450.35 & 1.12 & 9.05 \\
		SGHMC & 0.96 & 0.72 & 0.82 & 425.59 & 4.96 & 5.16 \\
		\hline
	\end{tabular}
	\caption{\label{tab_results_HMC}\footnotesize{Performance assessment and summary statistics for the ABP2 using both profile and network data (Scenario 1). Time is given in seconds per 100 iterations using a standard laptop with 16GB of RAM and a 2.60GHz Intel Core i7 processor.}}
\end{table}

\section{Discussion}\label{sec_discussion}

We have proposed a novel approach for de-duplication that easily reconciles both profile and network data. We have also developed a new prior specification on the cluster assignments, the ABP, which is easy to implement, naturally satisfies the microclustering property, and also makes it  straightforward to incorporate prior believes about the linkage structure.
Our experiments show that our formulation is quite robust to prior specification and outperforms its competitors by substantially improving the accuracy of the posterior linkage, and as a consequence, the estimate of the population size as well. We have also considered stochastic gradient Hamiltonian Monte Carlo methods in order to speed up the de-duplication process maintaining reasonable levels of accuracy.

Our work opens several doors for future research. We could either add an extra hierarchy to model the size of the larger cluster $M$ in a way that microclustering is preserved or consider a different joint distribution for the corresponding allelic partition $\rv$. Lastly, it also may be worth considering other fast approximation techniques in the flavor of variational approximations (\citealp{saul-1996}, \citealp{jordan-1998}, \citealp{beal-2003}, \citealp{broderick-2014}). This would allow us to consider bigger datasets with even millions of records.

\bibliographystyle{apalike}
\bibliography{RL}

\appendix

\section{Computation for ER Model}\label{ap_RL1}

\subsection{Markov chain Monte Carlo}

Consider the MCMC algorithm presented in \cite{sosa2018record}. No further steps are required when the UP is considered as the prior distribution for $\xiv$. However, if the ABP2 is used, note that
$$
p(\xiv\mid\rest) \propto (I-2r_2)!\,2^{r_2}\, r_2!\,\binom{Q_2}{r_2}\theta_2^{r_2}(1-\theta_2)^{Q_2-r_2}
$$
in step 1, and to complete the sampler, we need to add the following step to the algorithm:
\begin{enumerate}
	\item[9.] Sample $\theta_2^{(s+1)}$ from
	$p(\theta_2\mid\rest) = \Bet\left(\theta_2\mid a_2 + r_2, b_2 + Q_2 - r_2\right)$.
\end{enumerate}

On the other hand, if the EPP is used, note that
$
p(\xiv\mid\rest) \propto \theta^N \prod_{n=1}^N\Ga(S_n)
$
in step 1, and to complete the sampler, we need to introduce an auxiliary variable $\eta$ such that
$$
p(\theta,\eta\mid\rest)\propto p(\theta)\,\theta^{N-1}(\te+I)\times\eta^\te(1-\eta)^{I-1}.
$$
By doing so, we need to add the following step to the algorithm:
\begin{enumerate}
	\item[9.] Sample $\theta^{(s+1)}$ from the two-component gamma mixture:
	$$
	p(\theta\mid\rest) = \eps\,\Gamd(\te\mid a_\te+N,b_\te-\log\eta) + (1-\eps)\Gamd(\te\mid a_\te+N-1,b_\te-\log\eta)
	$$
	where
	$
	\eps=\tfrac{a_\te+N-1}{I(b_\te-\log\eta) + a_\te+N-1}.
	$
	\item[10.] Sample $\eta^{(s+1)}$ from $p(\eta\mid\rest)=\Bet(\eta\mid\theta+1,I)$.
\end{enumerate}

\subsection{Stochastic gradient Hamiltonian Monte Carlo}

The following are the steps required to draw samples from $p(\tev\mid\mathrm{D})$ using a SGHMC algorithm:
\begin{enumerate}[1.]
	\item Draw $\tilde{\mathrm{D}}$ uniformly at random from $\mathrm{D}$.
	\item Re-sample the momentum $\rv^{(s)}$ from $\Nor(\zerov,\M)$.
	\item Set $(\tev_0,\rv_0)=(\tev^{(s)},\rv^{(s)})$.
	\item Simulate Hamiltonian dynamics:
	\begin{enumerate}[i.]
		\item $\rv_0 \leftarrow$ $\rv_0 - \tfrac{\eps}{2}\nabla\tilde{U}(\tev_0)$.
		\item For $i=1,\ldots,L$ do: $\tev_i\leftarrow\tev_{i-1} + \eps\M^{-1}\rv_{i-1}$ and $\rv_{i}\leftarrow\rv_{i-1}-\eps\nabla\tilde{U}(\tev_{i})$.
		\item $\rv_L\leftarrow\rv_L-\tfrac{\eps}{2}\nabla\tilde{U}(\tev_{L})$.
	\end{enumerate}
	\item Set $(\tev^*,\rv^*)=(\tev_{L},\rv_{L})$.
	\item Compute the acceptance probability $$a= \ex{H(\tev^*,\rv^*)-H(\tev^{(s)},\rv^{(s)})},$$ where $H(\tev,\rv) = \tilde{U}(\tev) + \tfrac12 \rv^T \M \rv$ is the Hamiltonian function.
	\item Let
	$$
	\tev^{(s+1)} =
	\left\{
	\begin{array}{ll}
	\tev^{*}, & \hbox{with probability $a$;} \\
	\tev^{(s)}, & \hbox{with probability $1-a$.}
	\end{array}
	\right.
	$$
\end{enumerate}

Now we take this algorithm to sample $\be$ and each $\uv_n$ , $n = 1,\ldots,N$, being $N=\max\{\xiv\}$ the total number of latent individuals, as follows:
\begin{enumerate}
	\item If $\tev=\beta$, then we have that:
	\begin{align*}
		U(\beta) &= - \sum_{i<i'} \le[y_{i,i'}\log\theta_{i,i'} + (1-y_{i,i'})\log(1-\theta_{i,i'}) \ri] - \tfrac{1}{\sqrt{2\pi\omega^2}}\,\ex{-\tfrac12\,\be^2},\\
		\nabla U(\beta) &= \sum_{i<i'}\le[ \expit\{-(2y_{i,i'}-1 )\eta_{i,i'} \}\ri] + \tfrac{\be}{\omega^2},
	\end{align*}
	where $\eta_{i,i'}=\be-\|\uv_{\xi_i} - \uv_{\xi_i'}\|$ and $\te_{i,i'}= \expit\{\eta_{i,i'}\}$.
	\item If $\tev=\uv_n$, then we have that:
	\begin{align*}
		U(\uv_n) &= - \sum_{i'\in R_i} \le[y_{i,i'}\log\theta_{i,i'} + (1-y_{i,i'})\log(1-\theta_{i,i'}) \ri]\\ 
		&\hspace{2cm} -(2\pi\sig^2)^{-K/2}\,\ex{-\tfrac{1}{2\sig^2}\,\uv_n^T\uv_n},\\
		\nabla U(\uv_n) &= \le[ \sum_{R_i} \expit\le\{-(2y_{i,i'}-1 )\eta_{i,i'}\ri\}\frac{u_{n,k} - u_{\xi_{i'},k}}{\|\uv_n - \uv_{\xi_{i'},k} \| } + \tfrac{u_{n,k}}{\sig^2} \ri],
	\end{align*}
	where $R_i=\{i\in[I]:\xi_i=n\}$
\end{enumerate}

\end{document}